\documentclass{article}
\pdfoutput=1
\usepackage{amsfonts,amsmath,amssymb,graphicx}

\def\be{\begin{eqnarray}}
\def\ee{\end{eqnarray}}
\def\nn{\nonumber}

\def\Tr{{\rm Tr}\,}

\def\l[{\phantom.[}

\hoffset= - 1.0in         % switch off for draft style
\begin{document}
\title{{\bf {Is the Standard Model saved asymptotically \\by conformal symmetry?} \vspace{.2cm}}
\author{{\bf A. Gorsky$^{a,b,}$}\footnote{gorsky@itep.ru}, {\bf A. Mironov$^{a,c,d,e,}$}\footnote{mironov@itep.ru; mironov@lpi.ru}, {\bf A. Morozov$^{a,d,e,}$}\thanks{morozov@itep.ru} \ and {\bf T.N. Tomaras$^{f,}$}\thanks{tomaras@physics.uoc.gr}}
\date{ }
}

\maketitle

\vspace{-6.0cm}

\begin{center}
\hfill FIAN/TD-12/14\\
\hfill IITP/TH-01/14\\
\hfill ITCP-2014-11\\
\hfill ITEP/TH-24/14\\
\end{center}

\vspace{4.2cm}

\begin{center}
$^a$ {\small {\it Institute for Information Transmission Problems, Moscow 127994, Russia}}\\
$^b$ {\small {\it Moscow Institute of Physics and Technology, Dolgoprudny 141700, Russia }}\\
$^c$ {\small {\it Lebedev Physics Institute, Moscow 119991, Russia}}\\
$^d$ {\small {\it ITEP, Moscow 117218, Russia}}\\
$^e$ {\small {\it Moscow Physical Engineering Institute, Moscow 115409, Russia }}\\
$^f$ {\small {\it Department of Physics and Institute of Theoretical and Computational Physics, \\
University of Crete, 71003, Heraklion, Greece}}
\end{center}

\vspace{1cm}

\begin{abstract}
It is pointed out that the top-quark and Higgs masses and the Higgs VEV satisfy with great accuracy the relations $4m_H^2=2m_T^2=v^2$, which are very special and reminiscent of analogous ones at Argyres - Douglas points with enhanced conformal symmetry.
Furthermore, the RG evolution of the corresponding Higgs self-interaction and Yukawa
couplings $\lambda(0)=\frac{1}{8}$ and $y(0)=1$ leads to the free-field stable point $\lambda(M_{Pl})= \dot\lambda(M_{Pl})=0$ in the pure scalar sector at the Planck scale,
also suggesting enhanced conformal symmetry. Thus, it is conceivable that the Standard Model is the low-energy limit of a distinct special theory with (super?) conformal symmetry
at the Planck scale. In the context of such a ``scenario" one may further speculate that the Higgs particle is
the Goldstone boson of (partly) spontaneously broken conformal symmetry. This would simultaneously resolve the hierarchy and Landau pole problems in the scalar sector and would provide a nearly flat potential
with two almost degenerate minima at the electroweak and Planck scales.
\end{abstract}

%\vspace{1cm}

\section{Introduction}

Scalar theory, unless it is free, suffers from two severe problems:
the Moscow zero (Landau pole) problem \cite{zcp},
well established in lattice calculations \cite{lat}
and constructive field theory \cite{cons}, and the hierarchy problem.
This could cast a dark shadow on the Standard Model (SM),
which depends crucially on the scalar Higgs field.
The most popular ways to avoid them, propose serious modifications
of the SM at the TeV regime, either by adding
super-partners to known elementary particles, or by
making some of them composite, or both.
However, increasing attention is received recently by an alternative paradigm \cite{FN}-\cite{NielBled}, according to which there can be no
new physics beyond the SM all the way up to or around the
Planck scale, that the above problems of the scalar sector are red herrings
and that the apparent fine-tuning of the Higgs potential
is in fact an inescapable consequence of its distinct form
in a healthy fundamental theory defined at Planck energies.

The main arguments in favor of this scenario are based on the very special values
of the Higgs and the top-quark masses $m_H$ and $m_T$,
or, equivalently, of the low-energy values of the most relevant to our discussion scalar self-coupling $\lambda(0)$,
the three gauge couplings $\vec g(0)\equiv(g_1(0), g_2(0), g_3(0))$ and the top Yukawa
coupling constant $y(0)$, which in the conventional approaches
are considered ``accidental coincidences", while in the alternative
one, very important evidence. Specifically, this ``scenario" builds upon the following {\it experimental} facts and aims at providing alternative resolutions of the corresponding puzzles:

\bigskip

{\bf Fact 1:} The values $\lambda(0), \vec g(0), y(0)$ are fine-tuned, so
that at the Planck scale, i.e. for $t=\log\mu^2\sim (0.5\div 1)\log M_{Pl}$ one obtains simultaneously
\be
& &\dot\lambda(M_{Pl}) = 0 \nonumber \\
& & \nonumber \\
& &\lambda(M_{Pl}) = 0
\label{laladot}
\ee

{\it Puzzle 1:}
This seems to suggest that Nature started at the Planck level at a very distinguished
point, where $\lambda$ is stable and vanishing (free scalar theory),
and after that the RG evolution, mainly due to the evolution of the gauge couplings,
which were not stable at $M_{Pl}$, brought the scalar field to its present state
with a very concrete potential.

Reverting the statement, is the $\lambda\phi^4$ sector of the standard model
fine-tuned to be ``asymptotically secure",
instead of exhibiting unhealthy Landau pole behavior?

\bigskip

{\bf Fact 2} (perhaps, related to 1):
According to \cite{pot,pot1}: it seems that
the values $\lambda(0), \vec g(0), y(0)$ are fine-tuned so,
that the effective potential for the scalar field has
in addition to the SM Higgs vacuum expectation value $<\!\phi\!>=v$ another local minimum
at $<\!\phi\!> \approx M_{Pl}$ and {\it nearly degenerate} with the standard one. Perhaps, our minimum at $v$ is even slightly  metastable, since the SM parameters may be lying in a very narrow
metastability region.

{\it Puzzle 2:} Does this form of the effective potential, which seems quite special, suggest something important about the fundamental theory of Nature, or is it just a coincidence?

\bigskip

{\bf Facts 3 \& 4 (two relations):}
\noindent
It is an experimental fact that the Higgs mass, the top-quark mass and $v$ satisfy, with miraculous accuracy, the
relations
\be
4m_H^2 = 2m_T^2 = v^2
\label{ADrels}
\ee
i.e. there seems to be a clear conspiracy between the Higgs, the top-quark and the W/Z-boson masses.
More precisely
\be
{\sqrt{2}m_T\over v}=0.9956\pm .0044\\
{\sqrt{2}m_H\over m_T}=1.0252\pm .0073\nn
\ee
These are the pole (hence, not running) masses, and the Higgs field vacuum expectation value
is defined from the value of the Fermi constant:
\be\label{masses}
m_H=125.66\pm 0.34 \hbox{ GeV}\,, \;\;\;
m_T=173.34\pm 0.76\hbox{ GeV}\,, \;\;\;
v={1\over 2^{1/4}\sqrt{G_F}}=246.21817\pm 0.00006\hbox{ GeV}
\ee
Using (\ref{masses}), one obtains for the Yukawa ($y$) and Higgs self-couplings ($\lambda$)
\be
 y=1 & & \left(m_T=\frac{v}{\sqrt{2}}\right) \nn \\
\lambda = \frac{1}{8} & & \left(m_H = \frac{v}{2} \right)
\label{ylambda}
\ee

{\it Puzzles 3 \& 4:}
Is it possible that these special values of the couplings and the corresponding mass relations point to some hidden symmetry underlying the SM, which should further enhance a conformal-like symmetry at
the Planck scale, that is strongly suggested by (\ref{laladot})? What this symmetry could be? Have we ever before encountered a similar situation? We will point out in Section 4, that such relations are reminiscent of the Argyres-Douglas point known to exist in certain theories with enhanced symmetry. In that context, such mass relations are consequences of the symmetries of the theory and should be stable against RG flow. In this connection, the following is a very welcome additional fact.

\bigskip

{\bf Fact 5:} The difference $\xi=|\frac{1}{8}y^2 - \lambda|<0.05$ remains small
all along the RG-evolution region, so that the Argyres-Douglas like relation
is RG-stable with relatively satisfactory accuracy.

{\it Puzzle 5:} However, this statement is  sensitive to the exact value of the top-quark mass
(which is so far obtained with good accuracy only by combining the results of 4 collaborations \cite{mT}).
Stability of the above difference gets especially well pronounced (see Fig. 3 below),
if the parameters of the SM are chosen so that (\ref{laladot}) are {\it exact},
as {\it expected} in the context of an alternative paradigm speculated here.
Does this adjustment really take place, when improved by higher-loop corrections
and more precise measurements?

Assuming that it does, this choice of the SM parameters leads to another interesting bonus, namely to the:

\bigskip

{\bf Fact 6:} For the values of the parameters of the Standard Model which lead to the relations (\ref{laladot}), the 1-loop effective potential has a second almost degenerate minimum at a field value practically equal to the Planck scale (see Fig. 4 below).

{\it Puzzle 6:}
Thus, the Planck scale, which is  not present
in the lagrangian  of the Standard Model, is nevertheless hidden in the actual values of its parameters and the conjectured property (\ref{laladot}) of the fundamental theory at $M_{Pl}$.

\bigskip

All these puzzling facts seem to imply
that the parameters of the Standard Model are not at all accidental.
Instead, they may be fully determined by an assumption, that the Standard Model
is a low energy limit of a very special fundamental theory defined naturally at the Planck scale, which is the next fundamental threshold in particle physics.
Moreover, these relations imply that there is some
{\it additional symmetry}, which underlies the Standard Model and the deeper fundamental theory.
This symmetry should automatically protect the vacuum expectation value of the Higgs field
(in order to protect relations like (\ref{ylambda})) and, hence, solve the hierarchy
problem (in the spirit of \cite{Bard,HP}). Clearly, one could not hope for more, but unfortunately, we cannot be more concrete at this stage.

In the rest of this paper we will elaborate briefly on the above facts and speculate about the nature of a theory in the framework of the less conventional scenario sketched here.

\section{RG flow to (or from) a very special UV-point}

\subsection{RG flow in the Standard Model}

In Fig. 1 we plot the curves describing the one-loop RG evolution
of the five couplings of the Standard Model (they are actually the same as those in \cite{CEQ,IRS}).
Our
notation and initial values of coupling constants coincide with \cite{pot1}.
Nowadays these results are enhanced to include two and three-loop
corrections \cite{pot,pot1}, but they only improve the level of the fine-tuning apparent already at one-loop.

\begin{figure}
\begin{center}
\includegraphics[scale=0.2]{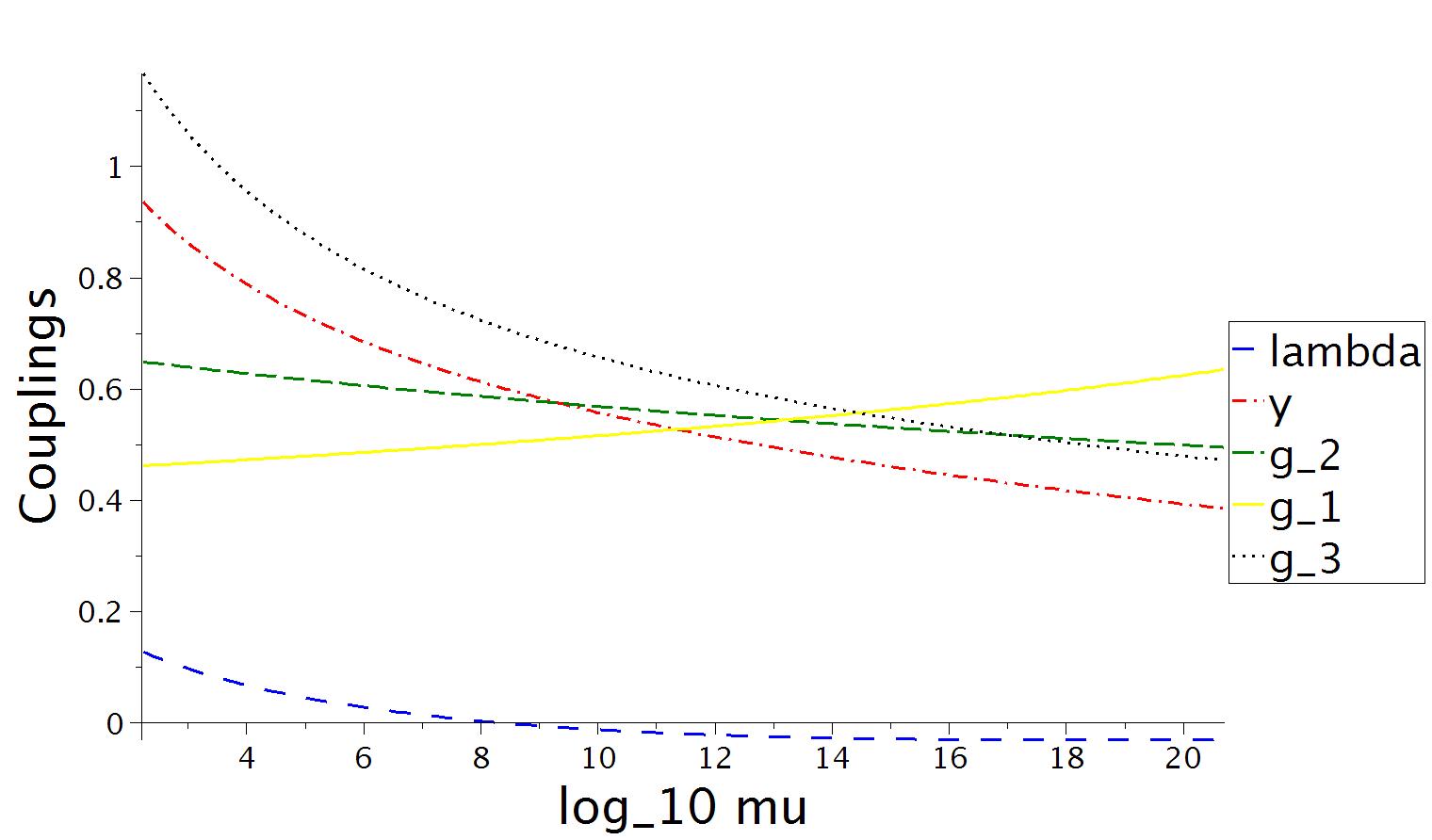}
\end{center}
\label{Fflows}
\caption{{\footnotesize RG flow of the coupling constants in one-loop approximation as a function of $\log_{10}\mu$ with the RG scale $\mu$ expressed
in GeV. Notice the well-known fact of $g_1$, $g_2$ and $g_3$ ``unification" at around $10^{15}$ GeV.
Notice also that the asymptotic behavior of the
Higgs self-coupling is in good agreement with (\ref{laladot}), given that the initial low energy values used are
the experimental central values of the couplings.
}}
\end{figure}

Thus one can see that the actual values of $\lambda(0)$ and $y(0)$ in particular
are such that there is a region with the properties (\ref{laladot})
at approximately $\log_{10} \mu \ge 8.3$. With three loop accuracy \cite{pot,pot1} it is
shifted only slightly to $\log_{10}\mu\ge 8.5$, with even a bit closer to zero value of $\lambda (M_{Pl})$.
This means that the one-loop approximation is quite reliable and {\it we can recover the Standard Model at low energies,
starting from the theory with this very special property at the Planck scale}.

Schematically, the one-loop RG equations of the SM have the well-known form
(of the three gauge couplings only $g_3\equiv g$ is kept, being the most important one):
\be
\frac{d}{dt}\,g^2 = -\beta g^4,\nn \\
\frac{d}{dt}\, y^2 = \alpha y^4 - \gamma g^2y^2,\nn \\
\frac{d}{dt} \,\lambda = \lambda(a\lambda+by^2-cg^2)-\mu y^4+\nu g^4
\label{rgeqns}
\ee
Two remarks are in order about these equations: First, the {\it signs} of the various terms of the $\beta-$functions are, of course, not accidental, reflecting basic properties of the SM, e.g.
\be
\begin{array}{cl}
\beta>0 & \text{asymptotic freedom}\\
\alpha/ \gamma > 0 & \text{attraction/repulsion  in scalar/vector exchange} \\
a>0 & \text{Landau pole for scalar self-coupling} \\
\alpha >0 & \text{Landau pole for the Yukawa coupling}
\end{array}
\ee
However, $\alpha>0$ does not necessarily imply the existence of a Landau pole in the Yukawa coupling. Surprisingly,
this depends not only on the coefficients of the differential equations, but also on the initial values. Indeed,
the first two equations do not depend  on $\lambda$,
and their solution is
\be
\frac{1}{g^2}&=&\frac{1}{g_0^2} + \beta t, \nn \\
\frac{1}{y^2} &=& \frac{\alpha}{\gamma-\beta}\, \frac{1}{g^2}
+ C \left(\frac{1}{g^2}\right)^{\gamma/\beta}
\ee
In practice $\gamma/\beta>1$ and $C$ is defined by the initial condition on the gauge and Yukawa couplings. The presence or absence of the Landau pole in $y(t)$ depends on the sign of $C$.
Finally, one can now solve the third equation, for $\lambda$ using the above solutions for $g(t)$ and $y(t)$
to complete the RG-flows.

Second, note that the system (\ref{rgeqns}) has a {\it triangular} property,
allowing solutions to avoid chaotic behavior that could, in principle, lead to conclusions
very different \cite{cRG} from the one discussed here. Thus,
this triangular structure of (\ref{rgeqns}) can by itself serve as an argument in support of the idea that
the above set of RG equations is very special and encodes important properties of the SM.

\subsection{Asymptotically secure Higgs}

If one wants to ``secure" the UV behavior of the scalar sector
at the Planck scale in the way explained in the Introduction, then
at $t=\log M\approx \log M_{Pl}$ one should require that $\lambda=0$ and $\dot\lambda=0$
\footnote{We call this {\it asymptotically secure} situation and not ``asymptotically safe", because the latter usually refers to a non-trivial
fixed point, while in our case the coupling is supposed to vanish. An example of the asymptotically safe Higgs theory can be found in
\cite{Litim}.}. This requirement fixes the initial low energy value $\lambda(0)$ of the Higgs self-coupling.
Indeed, given the RG equation for $\lambda$
\be
\dot\lambda = a\lambda^2 + \lambda f(\vec g,y) + h(\vec g,y)
\label{ladot}
\ee
and the evolution laws $\vec g(t), y(t)$
one may find the scale $\mu$ at which $\lambda=0=\dot\lambda$ from the equation
$
h\Big(\vec g(\mu),y(\mu)\Big) = 0\,,
$
and then use (\ref{ladot}) to solve for $\lambda(t)$
with
$
\lambda(\mu)=0\,.
$
This gives the asymptotically secure fine-tuned value for $\lambda(0)$ at low energies. The fact that this procedure, when applied to the full one-loop RG equations of the SM, gives $\mu\simeq M_{Pl}$ and for $\lambda(0)$ almost precisely the measured value of the Higgs coupling at the TeV scale (within the experimental error bar of one standard deviation), cannot in our opinion be considered as plain coincidence.

\section{Effective potential}

The running coupling $\lambda(t)$ is also relevant to the computation of the (RG improved) Coleman-Weinberg effective
potential \cite{EP} for $\phi$.
At one loop this effective potential is just \footnote{For large values of $\phi$ and with the quadratic divergences fine-tuned away, this is a very good approximation of the SM one loop effective potential.}
\be
V_{eff}(\phi)= \lambda(\phi)\, \phi^4
\label{Veff}
\ee
Normally, the zero of the beta-function of $\lambda$ means nothing special for the effective potential.
However, things are very different at a point like (\ref{laladot}), since at such a point $V_{eff}$ in (\ref{Veff}) has a {\it minimum}.
Furthermore, this minimum is especially spectacular, because it occurs at large $\phi\sim \mu \sim M_{Pl}$, where the classical potential $\lambda(0)\phi^4$ is extremely large.
This simple observation has recently been strengthened by a detailed analysis of the Standard Model \cite{pot,pot1}, which takes into account higher loop corrections.

First, let us consider the possibility that the values of $\lambda(0)$ and $y(0)$ are such that relations (\ref{laladot}) are satisfied {\it exactly}.
With these parameters the self-coupling $\lambda$ behaves as in Fig. 2, and the shape of the effective potential implied by the Standard Model is shown in Fig. 4.
\begin{figure}
\begin{center}
\includegraphics[scale=0.2]{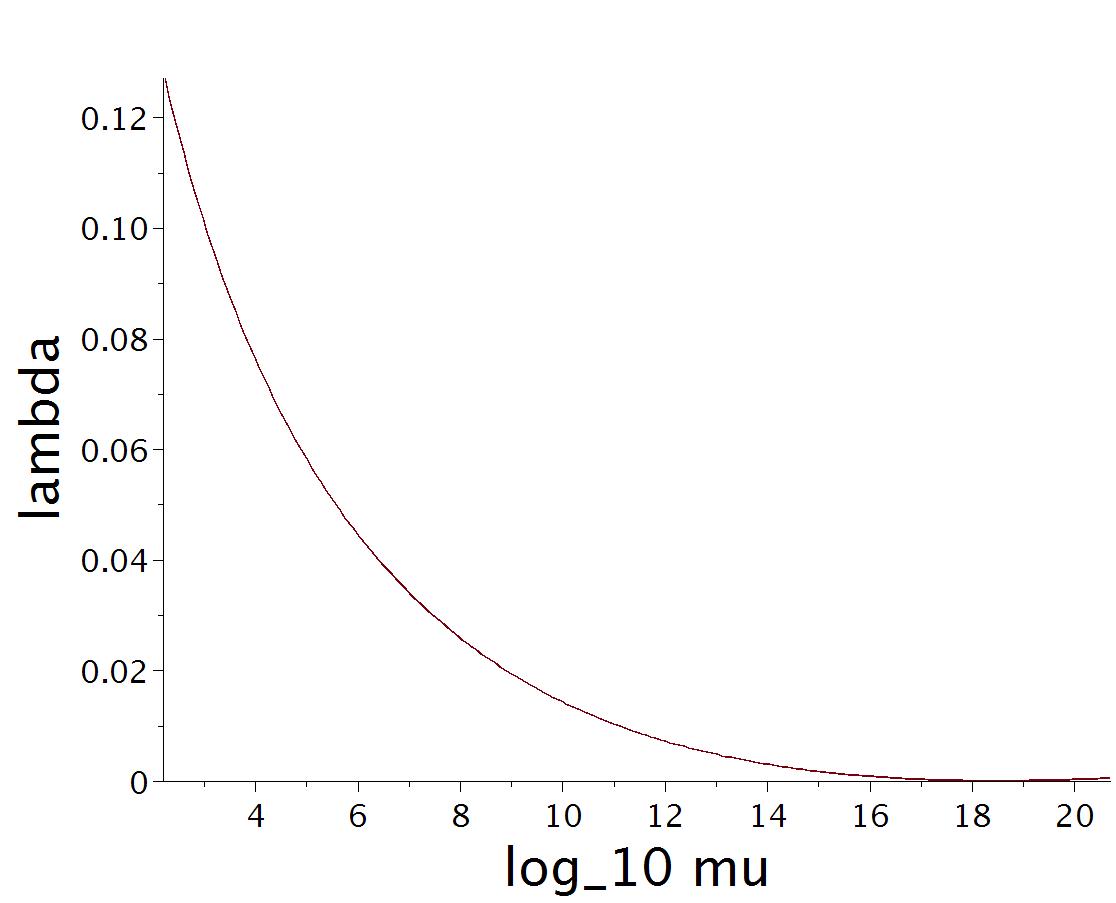}
\end{center}
\label{Flambda}
\caption{{\footnotesize The RG flow of the self-coupling $\lambda(\mu)$
in one-loop approximation as a function of $\log_{10}\mu$ with the RG scale $\mu$ expressed in GeV. The ``ideal" value of $y(m_T)$ that leads to relation (\ref{laladot}) was used. The Planck scale is obtained automatically as the ``touching point" of the curve to the abscissa axis. It is also instructive to look at the accuracy of the fine-tuning of $\lambda(\mu)$ and $y(\mu)$, needed to fit that special
point, at higher energies, see Fig. 3.} }
\end{figure}
Note that $\lambda$ touches zero just at the Planck scale, as
does the effective potential, so that its second minimum is also at the Planck scale\footnote{This position of minimum is, however,
gauge dependent, see \cite{LN}.}. Still, this is not the result of a careful fine-tuning. It is obtained simply by choosing the ratio of the Higgs to top masses so that $\lambda(t)$ just touches the horizontal axis and with the other SM parameters (e.g. the gauge couplings) taken from experiment. To within one standard deviation the values of these masses satisfy this requirement, i.e. lead to $\lambda(t)$ which touches the axis, and miraculously, give the extra bonus that the ``touching point" is obtained automatically at the Planck scale. Thus, to summarize,
it seems that within one standard deviation the parameters of the SM are such that relations (\ref{laladot}) are satisfied and the special behavior of $\lambda(t)$ and $V_{eff}(\phi)$ given above is obtained, with the Planck scale arising automatically.

\begin{figure}
\begin{center}
\includegraphics[scale=0.2]{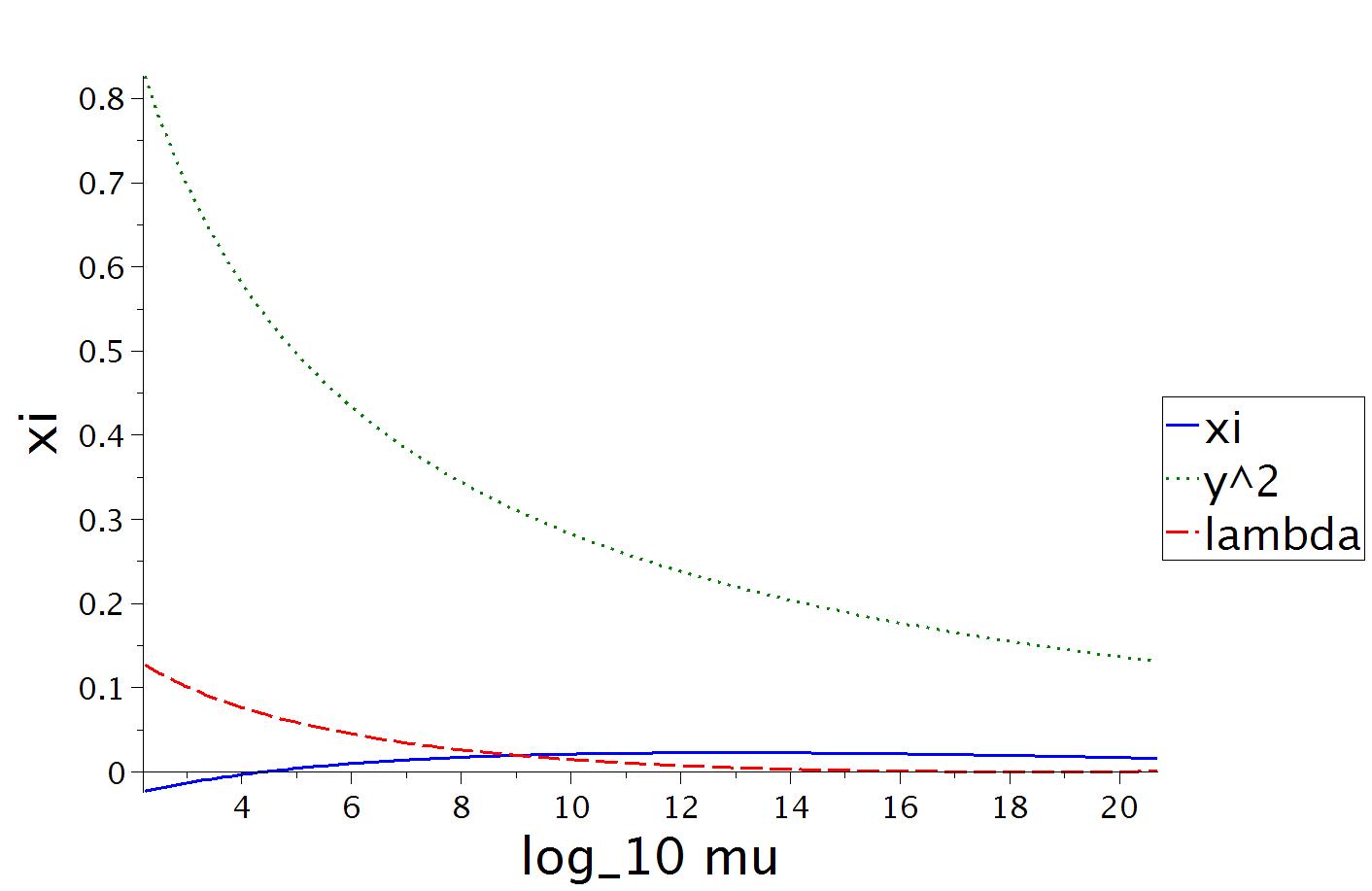}
\end{center}
\label{Fxi}
\caption{\footnotesize {Illustration of the Fact  5. The difference $\xi\equiv y^2(\mu)/8-\lambda(\mu)$
looks almost RG stable and
is much smaller than the values of $y$ and $\lambda$ themselves. The picture is the one-loop approximation, and for ``ideal"
values of parameters for which (\ref{laladot}) is exact.}}
\end{figure}

\begin{figure}
\begin{center}
\includegraphics[scale=0.2]{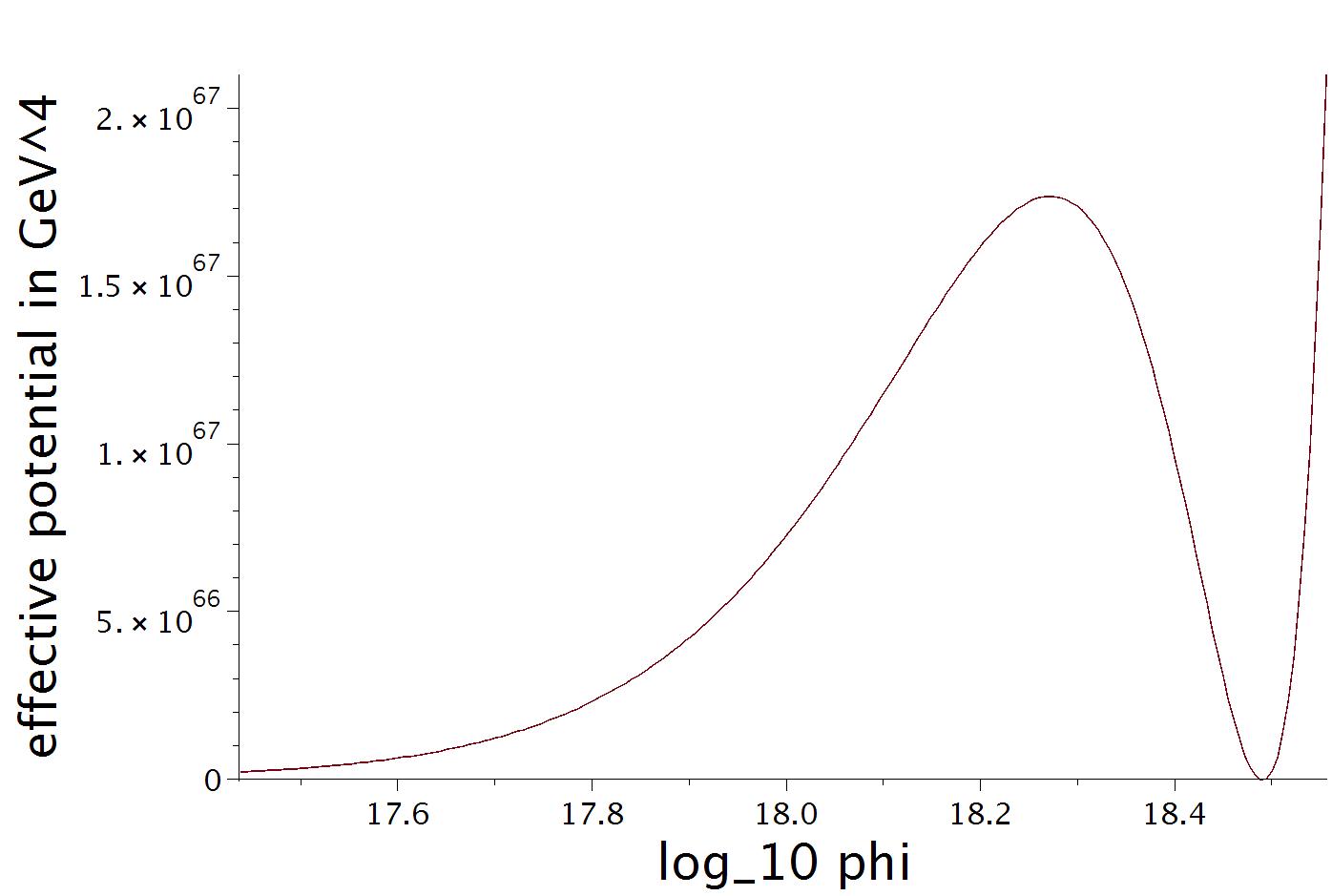}
\end{center}
\label{FVeff}
\caption{{\footnotesize
The improved one-loop effective potential for the value of $y(m_Y)$ that leads
to relation (\ref{laladot}) (see Fig. 2), as a function of $\log_{10}\phi$ with the field $\phi$
expressed in GeV. Note that, when the small uncertainty of the Standard Model parameters
is fixed, so that (1) is exactly satisfied, the second minimum of the potential is located
at the Planck scale. This illustrates Fact 6 of the Introduction. Note also that the potential
barrier is rather low, being seven orders of magnitude lower than its natural value $M_{Pl}^4$.
}}
\end{figure}

The Standard Model minimum, which is very close to the origin, is very shallow as compared with the height of the barrier in Fig. 4.
The location of the second minimum depends strongly on the parameters of the Standard Model. For the central values of the experimental SM parameters, the second minimum is deeper than the SM one, and may be located at energies even somewhat
larger than the Plank mass. However, the barrier is still high enough to guarantee that the metastable Standard Model vacuum has a lifetime much much longer than the age of the Universe \cite{lt, pot}.

Furthermore, the flatness of the potential in Fig. 4, the barrier of which is about seven orders of magnitude lower than its ``natural" scale $M_{Pl}^4$, fits nicely with the slow-roll requirement (${V''(\phi)/ V(\phi)}\ll M_{Pl}^{-2}$ and
${V'(\phi)/ V(\phi)}\ll M_{Pl}^{-1}$ \cite{GR}) of scalar fields in inflation models and has inspired several authors to investigate the possibility that the Higgs itself plays the role of the inflaton in such models \cite{inf=Higgs}. More recently this possibility has been studied e.g. in \cite{ASinf=Higfirst} and
\cite{ASinf=Higlast}.

Finally, it should be pointed out here, that the form of the potential (\ref{Veff}) restricted to the scalar sector is to leading order identical to the one obtained by Migdal and Shifman in \cite{MS, EL} to describe the low energy dynamics of the dilaton field, a Goldstone-like field which arises as a consequence of the spontaneous breaking of the conformal invariance in pure gluodynamics. The effective lagrangian of the dilaton with this potential was constructed on general grounds and is exactly the one which guarantees the validity of the corresponding Ward identities. Furthermore, an analysis within the Standard Model of the Higgs particle as a dilaton has been performed in \cite{EO}.

\section{Argyres-Douglas points with enhanced symmetry
\label{ADsec} }

If one asks whether there is any known situation,
when {\it a conformal symmetry emerges in some sector of the theory at a given
ratio of scalar/fermion masses},
an immediate answer is the Argyres-Douglas point.
The exact situation differs there from the Standard Model in many respects:
the theory is supersymmetric (originally it was ${\cal N}=2$ SUSY,
but actually ${\cal N}=1$ is enough, see, however, \cite{N=2SM}), the Higgs is hence in the adjoint representation,
it emerges in the infrared rather than in the ultraviolet region.
Still, it illustrates the main fact:
{\it emergence  of an extra symmetry in one sector of the theory
at some energies can be related to mysterious numerical relations observed in
another sector.}
Further studies can easily make the analogy much stronger.
Therefore, we briefly remind that old story.

\subsection{AD point in SUSY chromodynamics}

The low-energy sector of ${\cal N}=2$ SUSY theory is described by the
Seiberg-Witten (SW) theory \cite{SW}, where everything is encoded \cite{GKMMM}
in terms of a $0+1$-dimensional integrable system, associated with a
peculiar family of spectral curves $\Sigma$.
In particular, masses of the BPS states are given by periods of the
SW differential $d\theta=pdq$.
Whenever a non-contractible cycle on the Riemann surface shrinks to zero,
a BPS state becomes massless.
This happens at particular points (hypersurfaces) in the moduli space of
SW curves, i.e.   at  special values of vacuum average of the scalar
(adjoint Higgs) field.
In general at such points there is a singularity in the moduli space
but no any additional symmetry.

At Argyres-Douglas (AD) points \cite{AD},  {\it two}
cycles simultaneously shrink to zero.
At such points of the moduli space   {\it pairs} of massless BPS fields appear,
which are mutually non-local, and the Coulomb branch gets
described by a very interesting non-trivial
conformal theory \cite{confAD}.

\bigskip

\begin{picture}(200,200)(-140,-30)
\qbezier(0,0)(100,30)(200,10)
\qbezier(0,0)(10,30)(-30,60)
\qbezier(-30,60)(100,50)(180,80)
\qbezier(200,10)(170,30)(180,80)
\qbezier(20,15)(100,40)(170,70)
\qbezier(30,50)(100,40)(140,25)
\put(40,45){\line(-1,2){40}}
\put(155,60){\line(1,2){40}}
\put(92,40){\line(-1,2){30}}
\put(92,40){\line(1,2){30}}
\qbezier(62,100)(95,120)(122,100)
\put(151,43){\mbox{{\footnotesize moduli space,}}}
\put(149,35){\mbox{{\footnotesize  spanned by v.e.v. $<\phi>$   }}}
\put(168,27){\mbox{{\footnotesize   and hypermultiplet mass $\ m$}}}
\put(10,-20){\mbox{{\footnotesize the line $\oint_A d\theta = 0$}}}
\put(-70,-31){\mbox{{\footnotesize (additional massless BPS field of electric type)}}}
\put(140,-12){\mbox{{\footnotesize the line $\oint_B d\theta = 0$}}}
\put(130,-24){\mbox{{\footnotesize (additional massless BPS field of magnetic type)}}}
\put(80,25){\mbox{{\footnotesize AD point}}}
\put(100,160){\mbox{{\footnotesize additional}}}
\put(90,150){\mbox{{\footnotesize massless BPS fields}}}
\put(60,128){\mbox{{\footnotesize theory with extra}}}
\put(50,120){\mbox{{\footnotesize conformal symmetry }}}
\put(80,150){\vector(-2,-1){65}}
\put(155,145){\vector(2,-1){29}}
\put(120,115){\vector(-1,-1){15}}
\put(40,-10){\vector(1,2){17}}
\put(140,0){\vector(-1,2){12}}
\put(92,39){\circle*{4}}
\end{picture}

\bigskip

Concrete formulas behind this description in the simplest possible case,
the $SU(2)$ theory with one fundamental matter hypermultiplet
are as follows:
\be
\begin{array}{rl}
\hbox{The theory:}&\text{${\cal N}=2$ gauge supermultiplet + fundamental matter hypermultiplet}
\\
&
\\
\hbox{The family of curves:}& y^2=(x^2-u)^2-\Lambda^3 (x-m_T)
\end{array}
\ee
where the parameter $\Lambda$ is associated with $\Lambda_{QCD}$, $m_T$ with mass of the fundamental hypermultiplet
and the moduli of the curve
$u=<\hbox{Tr}\Phi^2>$ is related to the vacuum expectation $v$ of the adjoint scalar field (from the gauge supermultiplet) by
Seiberg-Witten theory\footnote{In terms of integrable systems, this model is \cite{GMMM}
a degeneration of the XXX spin chain at 2 sites, and $u$ is the Hamiltonian of the degenerated spin chain and $v$ the action
variable. The Hamiltonian interpretation of the AD points has been discussed recently in \cite{gorskymilekhin}.}. At large $v$: $u=1/2v^2$.

The curve describes the torus and is a Riemann surface with 4 ramification points, i.e. two independent cycles $A$ and $B$.
At the AD point in the moduli space,
\be
u={3\over 4}\Lambda^2\nn\\
m_T^2={3\over 4}u
\ee
the three of these ramification points merge, and the two cycles degenerate. This leads to emerging simultaneously massless
monopoles and charged states from the hypermultiplet and a non-trivial superconformal theory.

Now note that the supersymmetry requires that the superpotenital is of the form
\be
m_T\tilde\Psi\Psi-{1\over\sqrt{2}}\tilde\Psi\Phi\Psi+h.c.
\ee
After spontaneous breaking the symmetry, the two components of the hypermultiplet have masses $m_{\pm}=m_T\pm v/\sqrt{2}$.
In order to have conformal invariance, i.e. massless quark one requires
\be
m_T={v\over\sqrt{2}}
\ee
Thus, after breaking the symmetry one of the hypermuliplet components becomes massless, while
the other one gets mass $2m_T$, and this is the corollary of the supersymmetry (unit Yukawa constant) and conformal invariance.

\subsection{Breaking SUSY from ${\cal N}=2$ to ${\cal N}=1$}

One can explicitly break ${\cal N}=2$ supersymmetry down to ${\cal N}=1$
by adding a superpotential
\be
{\cal W} = \sum_k g_k \Tr \Phi^k+\hbox{fermionic interactions}
\ee
In the $SU(2)$-case described above the superpotential contain the massive term of the adjoint scalar field
$M\hbox{Tr}\Phi^2$ and fermionic interactions.

The singular points of the Coulomb moduli space  upon the perturbation becomes vacua in $N=1$ theory where the AD point is the point where two vacua collide. It was shown \cite{GVY} that both the monopole and charge
condensates vanish at this point, and the theory remains superconformal even after the strong breaking of $N=2$ to $N=1$.
Therefore physically the critical behaviour at AD point corresponds to the deconfinement phase transition.

Note that the condensates in this theory can be explicitly described within the technique developed in \cite{Cacha,DV}. At the
AD point they turned out to be related with parameters of the superpotential by simple relations. For instance, in the described
above case, $v=\sqrt{2}m_T$ and $u=m_T^2+{3/16\Lambda^2}$ (i.e. in the limit of large $m_T$ still $v^2=2u$).

The AD points and domains have been studied in various examples in SW theory,
with different field content \cite{AD,confAD,Cacha,DV,ADlast}.

\section{Related ideas}

\subsection{The multicriticality  principle by Froggatt and Nielsen}

Perhaps, the first who attempted to make a strong case against an intermediate energy scale between Fermi and Planck on the basis of RG properties were C. Froggatt and H.Nielsen \cite{FN}. They used earlier results of \cite{CEQ}, where the  requirement of positivity of the scalar potential led to {\it constraints} on the Higgs mass. Instead, Froggatt and Nielsen demanded that the minima of the scalar potential be exactly degenerate and {\it predicted} the correct value of the Higgs mass, seventeen years before it was finally announced at CERN \cite{cernann}.

To {\it justify} from first principles {\it why} Nature chooses this degeneracy, it was noted
in \cite{FN} that if in a multiphase thermodynamic system {\it extensive} parameters (like energy,
number of particles and volume) are fixed  instead of {\it intensive} ones
(like temperature, chemical potential and pressure), the system is automatically driven to the multicritical
(say, triple) point, where all the phases coexist in thermodynamic equilibrium, so that the intensive parameters are also {\it fixed}.
Taking for system the multiverse and for intensive parameters the shape of the effective potential, makes the
``multicriticality principle", i.e. that the possible vacua of the effective potential should be degenerate,
somehow justified and, perhaps, even attractive.
It differs significantly from the anthropic principle \cite{ant}, since it relies on ordinary fundamental physics without {\it a posteriori} assumptions like existence of galaxies, planets, life and consciousness.

\subsection{Models with t-quark condensates}

A well known scenario in which the masses of the Higgs boson and the top-quark are related is based on the Nambu - Jona-Lasinio original ideas and is described in \cite{Bar,M}. A four-fermi interaction is added to the SM action and the formation of a top-quark condensate is assumed to form, with characteristic compositeness scale $\Lambda\sim 10^{15}\div 10^{19}$ GeV. The Higgs boson emerges as a scalar excitation over the condensate and the top-quark mass turns out to be around $m_T\sim 200$ GeV. Finally, the masses of the scalar (Higgs) excitation and the top-quark are shown to satisfy simple relations, like the so called Nambu relation $m_H=2m_T$, which, however, are model dependent.

The two basic features of this scenario that make contact with our discussion are:
(i) the huge difference between the particle masses $m_H$, $m_T$ and the compositeness scale $\Lambda$,
which implies that the theory is ``almost conformal", a feature shared by the ${\cal N}=1$ model discussed in Section 4 near the AD point; (ii) the initial condition used in \cite{Bar} for the renormalization group at the compositeness scale $\Lambda$ is the vanishing of the scalar self-coupling, which corresponds to one of the two conditions in (\ref{laladot}).

\subsection{Asymptotically safe gravity}

The idea of asymptotically safe theories, put forward by S. Weinberg \cite{Wein}, has not so far attracted the attention it deserves, with the exception of asymptotically safe {\it gravity}, which is relatively well studied primarily by M. Reuter \cite{ASgrav}.

This is a radical idea with today's standards, since it admits that there is no new physics
beyond the Standard Model even at the Planck scale or above it \footnote{If strings do not show up there, there is no obstacle to go to higher scales, only in string theory the regions above and below Planck mass are dual to each other.}.
In such a context, it is natural to unify the ideas of
asymptotically secure Higgs and asymptotically safe gravity, as was strongly advocated in \cite{Sha}.

\subsection{Brane interpretation}

Like any Yang-Mills theory, the Standard Model can be embedded in various
brane backgrounds and it is interesting to discuss their properties
from the perspective of the present paper.
In a brane picture all condensates and other moduli are interpreted as
distances and fluxes in extra dimensions.
One could speculate that the remarkable ``numerical coincidences", described in the Introduction,
are needed for the stability of the whole brane configuration in a wide range of energies or, equivalently, values of the radial RG coordinate.
Approximate ``flatness" of the Higgs potential could imply that the brane configuration is nearly BPS,
since the flatness of the potential requires
cancelation of the interaction between the corresponding branes.

Another possible source of
relations between parameters is the matching of theories
on the ``flavor" and ``color" branes.
The theories on these branes are essentially different
(for example, one is Abelian in the case of one flavor while the other is not),
however all physical phenomena should be equally well described in terms of both branes.
The familiar example of this phenomenon is the equivalent description of conventional QCD as a
theory on the flavor branes (chiral Lagrangian) or as a theory on the color branes (QCD Lagrangian).
Another example is the $2d/4d$ correspondence, when the $4d$ physics can be equivalently
described by the $2d$ theory at the non-Abelian string.
An interesting kind of matching condition is provided by the decoupling of a heavy flavor.
The conformal anomaly implies that the condensate of the fermion field disappears with
increase of its mass:
$m<\!\!\tilde\Psi \Psi\!\!>\  =\  <\!\!{\rm Tr}\, G^2\!\!>$. \
This relation turns out to be part of the stability condition
of the brane geometry \cite{gorer} and holds in all QCD-related
backgrounds.

If the Standard Model is indeed at the borderline of metastability,
an interesting question is to understand what becomes unstable in the brane picture.
In the well-controlled supersymmetric context the AD point lies at the marginal
stability line/surface, where unstable in the ${\cal N}=2$ case are BPS particles, but in the ${\cal N}=1$ case
unstable are instead the
extended objects - domain walls
\cite{GVY}.
It is much less clear, what would happen when supersymmetry is completely broken, but one could imagine that
the metastability of the Standard Model vacuum
reflects a metastability of the ``color brane" at an AD-like point in the parameter space.

\section{Discussion}

Usually the biggest obstacle to the idea that there is no new physics
in between the Fermi and the Planck scales is the hierarchy problem: one should  explain, {\it why} quadratic divergences do not generate a scalar mass of the Planck size (for recent discussion see \cite{Natur}).
Together with the similar cosmological constant problem
it clearly implies that power divergences should be ignored in the Standard Model.
Moreover, even supersymmetry does not help, because, being broken, it is not sufficient to explain the smallness
of cosmological constant. The idea of asymptotic safety also is not sufficient, because the fact that the theory is very nice in the ultraviolet does not guarantee that unwanted contributions will not be generated by the RG evolution.
Power divergences are automatically absent in dimensional regularization schemes, but it is unclear whether the possible existence of small extra dimensions could really help.
Whatever one thinks about this problem, it is
{\it phenomenologically} clear that  quadratic divergences need to
be ignored in the Standard Model, and this is widely recognized
in the literature: it is enough to mention that the RG-evolution plots in refs. \cite{CEQ}
(the early counterpart of our Fig. 1) and \cite{pot,pot1} included evolution of the mass term,
but only logarithmic corrections were taken into account and considered as relevant to ``real" physics.

As for {\it explanations}, the hope may be that the ``hidden symmetry", reflected
in relations such as (\ref{ylambda}), could provide a new tool for the resolution of the hierarchy problem, since the symmetry would protect these relations, in particular,
leaving no room for quadratic divergences.
In fact, though not sufficiently appreciated, the idea that the apparent conformal symmetry
of the Standard Model at the classical level could forbid the generation of quadratic
corrections at the quantum level, has been discussed in the literature -- it is best expressed
in \cite{Bard}, where even a concrete quantization scheme is suggested.
This idea is also studied in \cite{MS}, which we mentioned in Section 3, or very recently in \cite{nicolai}
and, in a context related with the neutrino mass mechanism, in \cite{Sm}.

We want to emphasize that these ideas get additional support from our Fact 1.
Usually, classical conformal symmetry of the Standard Model is broken softly
by mass terms and seriously by (logarithmic) quantum corrections, giving rise to
non-vanishing beta-functions.
Our Fact 1 implies that the only role of the beta-functions is to drive the theory away
from the UV point -- but exactly {\it there} approximate conformal symmetry is actually enhanced:
in the scalar sector the beta-function is vanishing and the interaction is also vanishing.
The theory looks even more conformal than one could expect.
And this is further supported by extreme flatness of effective potential
(from Fig. 4 it is clear that the height of the barrier is seven orders of
magnitude {\it lower} than the naive $M_{Pl}^4$, while the mass of the scalar mode at the
Planckian minimum is instead {\it higher} by many orders of magnitude than the naive $M_{Pl}$,
so that it can be actually ignored) -- and all this is just an {\it experimental fact!},
following from the well-established properties of the Standard Model itself,
with no reference to any kind of ``new physics", nothing to say about
quantum gravity and string theory: the Planck scale appears in Fig. 4
just from the study of RG evolution of the Standard Model itself(!).
The only assumption is to neglect the quadratic quantum corrections-- but given
not just the classical conformal symmetry of \cite{Bard}, but its further
enhancement by (\ref{laladot}) at the ``starting point" in the ultraviolet,
one can hardly be surprised that they {\it should} be neglected
in appropriate quantization scheme.
In our view, it is now a clear challenge for string theory or whatever is the UV completion of the Standard Model to make such a scheme {\it natural}.

As we mentioned, within ordinary quantum field theory one option would be to look
for a formulation, where the Higgs scalars are actually Goldstones of spontaneously
broken conformal symmetry, which get relatively small masses due to the explicit
breaking of this symmetry by beta-functions, as implied by the analogy with
a similar situation in \cite{MS}.
However, in this general review we prefer not to speculate further about particular realizations of this option.

\section{Conclusion}

Inspired by the old works of Froggatt-Nielsen-Takanishi \cite{FN,FNT} on one side,
and by the spectacular relations among the parameters of the Standard Model on the other,
we reviewed the evidence that the {\it Standard Model lies at a very special point of the parameter space}.
Namely, that it is {\it connected by the RG evolution to
a theory with enhanced (conformal-like) symmetry} at Planck energies,
where it is supposed to be mixed with quantum gravity and, perhaps, string theory.
If true, this implies the exciting possibility that the actual values of couplings, which may seem
fine tuned at our energies, may just reflect the fact that we are looking at the {\it low energy limit of an UV-healthy theory},
thus providing a kind of {\it refinement of the renormalisability
principle}. In other words, it is possible that the low-energy theory
is not only necessarily a gauge theory, but in addition
its scalar sector should be very special, just as a consequence of being a low-energy effective theory.
This option, if actually realized, would resolve at once many puzzles about the
Standard Model.

We emphasized also that the well-known
interconnected {\it facts} 1 \& 2 about the Standard Model,
are complemented by {\it the facts} 3, 4 \& 5.
We mentioned that these two seemingly unrelated properties, namely the
existence of an enhanced conformal-like symmetry at one scale (facts 1 \& 2)
and remarkably special numerical relations at another (facts 3 \& 4),
which in addition look RG stable (fact 5), may well be related with each other.
At least one example with similar properties is already known:
at the Argyres-Douglas point a conformal symmetry in the BPS sector
emerges at the very special points in the original moduli space of vacuum expectation values and couplings.
In the Standard Model a conformal symmetry (probably) emerges in the ultraviolet and not in the infrared, but this is rather
an advantage, because this {\it explains, why} we should wish to adjust the parameters of the moduli space to be at this special AD point.

\bigskip

To summarize,
\begin{itemize}
\item
Problems  of the Higgs sector (zero charge and hierarchy)
could be naturally resolved by treating it as a low-energy
limit of an especially nice theory at the Planck scale.
\item
That theory can be at least conformal,
or, perhaps, even superconformal invariant.
This not only seems to match nicely with expectations based on string theory,
but it also looks {\it phenomenologically} motivated by the actual features of the Standard Model.
\item
As a dream-like scenario,
the Higgs sector could actually emerge as a Goldstone one,
associated with spontaneous breaking of a high-energy
conformal invariance, and this could solve both the hierarchy
and the Landau pole problems.
\end{itemize}

\section*{Acknowledgements}

A.M. and A.M. would like to express their gratitude to the ITCP (formerly Institute of Plasma Physics) of the University of Crete for its hospitality and TNT wishes to thank the LPTENS for its hospitality during the late stages of this work. We are indebted to M.Voloshin for valuable discussions.
Our work is partly supported by grant
NSh-1500.2014.2 (A.M.'s), by RFBR  grants 13-02-00457 (A.Mir.), 13-02-00478 (A.Mor.), 12-02-00284 (A.G.),
by joint grants 13-02-91371-ST, 14-01-92691-Ind,
by the Brazil National Counsel of Scientific and
Technological Development (A.Mor.) and by PICS-12-02-91052 (A.G.).
The work of TNT was supported in
part by European Union's Seventh Framework Programme under the EU program ``Thales" MIS 375734 and the FP7-REGPOT-2012-2013-1 no 316165 and was also co-financed by the European Union (European Social Fund, ESF) and Greek national funds through the Operational Program ``Education and Lifelong Learning'' of the National Strategic Reference Framework (NSRF) under the  ``ARISTEIA'' Action.

\end{document}